\documentclass[preprint,preprintnumbers,aps,showpacs,superscriptaddress,citeautoscript,nofootinbib]{revtex4-1}

\usepackage[utf8]{inputenc}
\usepackage{amsmath}
\usepackage{graphicx}
\usepackage{amssymb}
\usepackage{times}
\usepackage{color}
\usepackage{float}
\usepackage{multirow}
\usepackage{subcaption}
\usepackage{braket}
\usepackage{hyperref}
\usepackage{makecell}

\usepackage[nolist]{acronym}
\begin{document}

\begin{acronym}
	\acro{SiC}{Silicon carbide}
	\acro{ZPL}{zero phonon line}
	\acro{TDM}{transition dipole moment}
	\acro{ZFS}{zero field splitting}
	\acro{EPR}{electron paramagnetic resonance}
	\acro{PL}{photoluminescence}
	\acro{ADAQ}{Automatic Defect Analysis and Qualification}
\end{acronym}

\title{Exhaustive characterization of modified Si vacancies in 4H-SiC}

\author{Joel Davidsson}
\email{joel.davidsson@liu.se}
\affiliation{Department of Physics, Chemistry and Biology, Link\"oping University, Link\"oping, Sweden}
  
\author{Rohit Babar}
\affiliation{Department of Physics, Chemistry and Biology, Link\"oping University, Link\"oping, Sweden}

\author{Danial Shafizadeh}
\affiliation{Department of Physics, Chemistry and Biology, Link\"oping University, Link\"oping, Sweden}

\author{Ivan G. Ivanov}
\affiliation{Department of Physics, Chemistry and Biology, Link\"oping University, Link\"oping, Sweden}

\author{Viktor Iv\'ady}
\affiliation{Department of Physics, Chemistry and Biology, Link\"oping University, Link\"oping, Sweden}
\affiliation{Max-Planck-Institut f\"{u}r Physik komplexer Systeme, Dresden, Germany}

\author{Rickard Armiento}
\affiliation{Department of Physics, Chemistry and Biology, Link\"oping University, Link\"oping, Sweden}

\author{Igor A. Abrikosov}
\affiliation{Department of Physics, Chemistry and Biology, Link\"oping University, Link\"oping, Sweden}

\begin{abstract}

The negatively charged silicon vacancy ($\mathrm{V_{Si}^-}$) in silicon carbide is a well-studied point defect for quantum applications.
At the same time, a closer inspection of ensemble photoluminescence and electron paramagnetic resonance measurements reveals an abundance of related but so far unidentified signals.
In this study, we search for defects in 4H-SiC that explain the above magneto-optical signals in a defect database generated by Automatic Defect Analysis and Qualification (ADAQ) workflows.
This search reveals only one class of atomic structures that exhibit silicon-vacancy-like properties in the data: a carbon antisite ($\mathrm{C_{Si}}$) within sub-nanometer distances from the silicon vacancy only slightly alters the latter without affecting the charge or spin state.
Such a perturbation is energetically bound.
We consider the formation of $\mathrm{V_{Si}^-+C_{Si}}$ up to 2 nm distance and report their zero phonon lines and zero field splitting values.
In addition, we perform high-resolution photoluminescence experiments in the silicon vacancy region and find an abundance of lines.
Comparing our computational and experimental results, several configurations show great agreement.
Our work demonstrates the effectiveness of a database with high-throughput results in the search for defects in quantum applications.

\end{abstract}

\maketitle


\section{Introduction}

\ac{SiC} has great potential for quantum information and sensing technologies due to the large-scale high-quality manufacturing and ease of integration into existing semiconductor devices.
The silicon vacancy in SiC~\cite{Torpo1999} is a well-studied defect with applications such as qubits, sensors, and single photon emitters~\cite{baranov2011silicon,Riedel2012Resonant,Soltamov12,Soykal17,siliconvacacnyreview,ivady2017identification}.
Several of these applications work at room-temperature~\cite{Soltamov12,siliconvacacnyreview}.
Many efforts are dedicated toward the controlled fabrication of such defects with various techniques.
Current endeavors focus on the fabrication of silicon vacancy arrays with both laser writing~\cite{Chen2019} and ion implantation~\cite{Pavunny2021,Babin2022,Wang2017array}.
To guide the ion implantation, several molecular dynamics studies of implanting hydrogen, helium, and silicon in \ac{SiC} have been carried out~\cite{MDsilvac2021,MD2silvac2021}.
All these efforts show that the silicon vacancy in SiC is a viable candidate for various quantum information and sensing technologies.

The negative charge state of the silicon vacancy has C$_{\mathrm{3v}}$ symmetry, a ground state with spin-3/2, and a rich many-body structure~\cite{soykal2016silicon}.
In 4H-SiC, there are only two non-equivalent positions for the silicon vacancy, denoted $h$ for hexagonal-like layer and $k$ for cubic-like layer.
However, \ac{EPR} experiments show at least six different signals related to a spin-3/2 defect.
Two of these signals ($T_{V1a}$ and $T_{V2a}$) have been accredited to the $h$ and $k$ configurations, respectively~\cite{ivady2017identification}.
The remaining four additional related signals ($T_{V1b}$ and $T_{V2b}$)~\cite{sorman2000silicon,Son_2019_Ligand} and ($R_1$ and $R_2$)\cite{Son_2019_Ligand} remain unidentified.

In addition to EPR, several \ac{PL} measurements have reported silicon-vacancy-like signals.
Five of these signals have been measured within a 19 MHz range, and there are additional small peaks whose origins are unknown~\cite{Nagy2019}.
Using high-resolution photoluminescence excitation, nine additional signals have been measured in 0.15 meV range without spectral diffusion~\cite{hunter2019}.
Banks et al. state that the strain and electric field perturbation are low, and these signals most likely correspond to other defects.
This is echoed by Ramsey et al., who state that the most likely source of such signals is nearby defects perturbing the silicon vacancy~\cite{Ramsay2020Relax}.
Furthermore, in an array of silicon vacancies created with ion implantation, 30\% of the measured spots showed even larger spectral drift, as large as 3 nm (5 meV), towards smaller energies~\cite{Pavunny2021}.
In addition, \ac{PL} signals more than 30 meV from the silicon vacancy, have been seen in other experiments~\cite{sorman2000silicon,nagy2018quantum,Fuchs2015,PhysRevB.62.16555,PhysRevB.62.16555,doi:10.1063/1.5045859}.
The most prominent line is around 1.412 eV (878 nm).
This line is also found in connection to measurements done for the L-lines---additional lines that span a 15 meV range next to the $h$ silicon vacancy and are tentatively suggested to be vibronic replicas of the silicon vacancy~\cite{L-lines}.

Before assigning the above signals to a novel defect, it is necessary to consider the role of other perturbative effects such as thermal vibration, strain, and surface termination.
Thermal vibrations can be eliminated as a potential source since many reported measurements are performed below 4.9 K.
Similarly, the signals cannot be attributed to phonon replicas since they appear at higher temperatures (15 K) at a larger energy difference (37 meV) from the \ac{ZPL}~\cite{Shang2020local}.
They could emerge from strain, where a large shift in \ac{ZPL} (26 meV) has been reported for 6H-SiC nanoparticles~\cite{Vasquez2020}, which corresponded to a 2.3\% basal strain~\cite{Udvarhelyi2020}.
This extreme shift observed in nanoparticles is probably not reasonable for bulk.
Udvarhelyi et al. found that the shift of the silicon vacancy is larger for axial strain than for basal strain and concluded that a shift of several meV would be possible in bulk.
It is also possible that the surface effect may shift the ZPL.
However, all these effects are small.
Some of the \ac{PL} shifts appear to be quite large, beyond the range of 30 meV.
Thus, these signals most likely correspond to an unknown defect.

Previous efforts to explore specific defect realizations that explain the EPR measurements have tested modifying the silicon vacancy with carbon vacancies along the c-axis~\cite{NatPhys14,Astakhov2016}.
This approach was extended to include all possible vacancies and antisites along the c-axis~\cite{Adam_mod_Si}.
Moreover, carbon antisites as second nearest neighbors slightly off the c-axis have also been considered.
As reported, in Ref.~\onlinecite{Adam_mod_Si}, one of the tested configurations could be responsible for the $R_1$ EPR signal.
However, no experimental agreement with PL has been found.
The above manually tested point defects models are limited in scope since they do not include impurities or defect clusters containing interstitials.
No thorough large-scale search for point defects that can explain the experimental observations has been conducted yet.

In this paper, we exhaustively show how a silicon vacancy modified by a carbon antisite is the only candidate among thousands of considered defects and characterize its different configurations.
Sec.~\ref{sec:ht} outlines how we search for point defects that explain the observations related to the silicon vacancy in data produced in high-throughput calculations~\cite{adaq_info,davidsson2021color,defect_hull} with \ac{ADAQ}~\cite{ADAQ} .
This search shows that silicon vacancies modified by carbon antisites are the only candidates among thousands of considered point defects.
The following section (Sec.~\ref{sec:mod}) presents this defect in detail and introduces a compact nomenclature.
Sec.~\ref{sec:res} presents both theoretical results carried out by additional manual calculations and experimental measurements carried out in this work to verify theoretical predictions.
Sec.~\ref{sec:dis} covers general trends for the modified vacancies, identification of configurations based on theoretical and experimental results, discussion about the lines closest to the isolated vacancies, and an outlook for future experiments.
Hence, we conclude that the experimental observations emanate from silicon vacancies modified by carbon antisites.

\section{High-Throughput Search}
\label{sec:ht}

\subsection{ADAQ Software and Data}

To narrow down the number of possible defects that fit the observed experimental data, we turn to the high-throughput data~\cite{adaq_info,davidsson2021color,defect_hull} produced by \ac{ADAQ}~\cite{ADAQ} that is implemented with the high-throughput toolkit \textit{httk}~\cite{Armiento2020}.
\ac{ADAQ} is a collection of automatic workflows designed to speed up the search for point defects.
It generates defects and calculates the most important properties such as total energy and \ac{ZPL} for one excitation for different charge and spin states in a screening workflow.
For detailed description of \ac{ADAQ}, see Ref.~\onlinecite{ADAQ}.
In brief, \ac{ADAQ} runs density functional theory~\cite{Hohenberg64,Kohn65} (DFT) calculations using the Vienna Ab initio Simulation Package (VASP)~\cite{VASP,VASP2} (v.5.4.4) with the semi-local exchange-correlation functional of Perdew, Burke, and Erzenerhof (PBE)~\cite{PBE}.
Due to the many VASP invocation for the different charge, spin, and excitation for point defects; \ac{ADAQ} is necessary to handle the vast amount of computations.

Previously, \ac{ADAQ} was employed on 4H-SiC and screened 8355 single and double intrinsic defects in 4H-SiC~\cite{davidsson2021color}.
The detailed report on the results collected in a database will be presented elsewhere~\cite{adaq_info,defect_hull}.
The defects were generated with the settings for \ac{ADAQ} to include double defects with a maximum distance of 3.5 Å.
These settings roughly correspond to point defect clusters with second nearest neighbors.
To keep track of the most stable defects, the concept of the defect hull is introduced~\cite{davidsson2021color} which consists of the point defects with the lowest formation energy for a given stoichiometry.
The defect hull is analogous to the convex hull of stability used to discuss the thermodynamical stability of bulk materials.

\subsection{Search for Silicon-Vacancy-Like Signals}

Here, we present two different ways of searching through the defect database generated by ADAQ for a defect that explains the experimental measurements.
First, the \ac{EPR} measurements show that the silicon-vacancy-like defect has spin-$3/2$.
By searching through the ground state results for the 8355 single and double intrinsic defects in 4H-SiC, 39 defect configurations with spin-$3/2$ are found.
Of these configurations, 24 contain a silicon vacancy. 
These can further be reduced by only including defects that are a maximum of 1 eV above the defect hull and have a positive binding energy.
The remaining 8 final entries consist of the 2 isolated silicon vacancy configurations and 6 configurations of a cluster containing a silicon vacancy with a carbon antisite at the second nearest neighbor.
Given these 4 search criteria that (i) the defect spin is limited to $S=3/2$, (ii) the defect included a silicon vacancy, (iii) the defect has a positive binding energy, and (vi) the defect is a maximum of 1 eV energy above the defect hull.
The only point defects that fit all of them are the silicon vacancy and a cluster consisting of a silicon vacancy with a carbon antisite at the second nearest neighbor.
Hence, we will refer to any combination of silicon vacancy and carbon antisite as modified vacancy in the rest of the paper.
The modified vacancy is on the defect hull for the stoichiometry of two missing silicons and one extra carbon.
Since \ac{ADAQ} is not limited to only defects along the c-axis, it allowed us to find additional configurations compared to previous searches~\cite{Adam_mod_Si}.

Second, the \ac{PL} results show similar \ac{ZPL}s in regions next to the silicon vacancies which are at 1.352 and 1.438 eV~\cite{ivady2017identification}.
Note that due to the use of the PBE functional, the \ac{ZPL}s are shifted down by 0.2 eV which is discussed in detail in Ref.~\onlinecite{ADAQ}.
Since all defects are calculated with the same level of theory, the search criteria for \ac{ZPL} are larger than 1 eV and smaller than 1.3 eV.
Combined with a maximum of 1 eV above the defect hull, these criteria give 9 final entries.
Here, 2 divacancy configurations emerge due to the wide \ac{ZPL} search range.
Disregarding them, we are left with 7 configurations, 1 silicon vacancy and 6 modified vacancies.
A \ac{ZPL} is missing for a silicon vacancy due to the settings in the screening workflow which is also discussed in detail in Ref.~\onlinecite{ADAQ}.
The estimated \ac{ZPL}s for the modified vacancies are 1.09, 1.13, 1.20, 1.22, 1.24, and 1.25 eV. 

Either way, when one searches for a defect with silicon-vacancy-like properties with results from \ac{EPR} or \ac{PL}, the modified vacancies are found in both cases.
With these search criteria, we exhaustively examine all defects in the database.
The modified vacancy is the only defect of the considered single and double intrinsic defects for 4H-SiC that can fit the experimental observations and is worth additional study.
It is important to underline that ADAQ workflows are implemented with the accuracy settings needed for high-throughput calculations.
This means significantly more accurate calculations are needed to characterize the modified vacancy properly.
Below we present these calculations and results. 

\section{Modified Vacancy---Silicon Vacancy Modified by Carbon Antisite}
\label{sec:mod}

\begin{figure*}[h!]
   \includegraphics[width=\columnwidth]{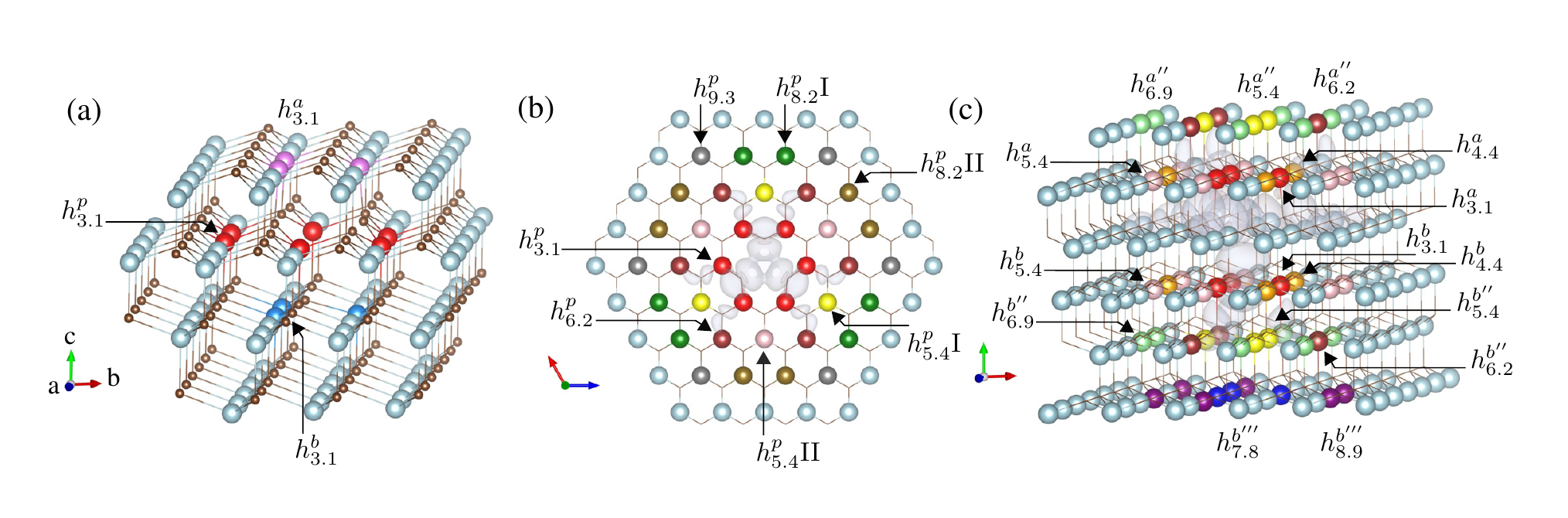}
	\caption{Possible sites of the carbon antisites for the $h$ silicon vacancy. The silicon atoms are colored grey and the carbons are colored brown. The equivalent atoms in C$_{3v}$ symmetry are colored with the same color in each subplot. The grey isosurface is the silicon vacancy spin density. a) the three symmetrically non-equivalent locations of the carbon antisite for the three closest configurations: $h_{3.1}^{a}$ (magenta),  $h_{3.1}^{p}$ (red) and $h_{3.1}^{b}$ (blue); b) a top view of the non-equivalent locations of the carbon antisite in the plane ($p$) with the silicon vacancy; c) the layers above ($a$) and below ($b$) of the silicon vacancy. Additional layers above and below are denoted by primes (') to indicate the layer distance. The color scheme is retained in subsequent plots to denote the distance between carbon antisite and silicon vacancy.}
	\label{fig:intro} 
\end{figure*}

Around each silicon vacancy configuration ($h$ and $k$) in 4H-SiC, there are three non-equivalent second nearest neighbor sites where a carbon antisite can be placed.
Hence, six different configurations exist for the closest modified vacancies, which are shown in Figure~\ref{fig:intro}a.

\ac{ADAQ} generated the closest configurations.
In addition, we also manually place the carbon antisite farther away from the silicon vacancy to study how the distance and orientation between the two defects affect the properties.
Figure~\ref{fig:intro}b and c show these configurations.
Here, the local site symmetry around the silicon vacancy is used to find the non-equivalent modified silicon vacancies.
The following nomenclature is used to keep track of the different configurations: $configuration_{distance}^{layer}index$.
$Configuration$ is either $h$ or $k$ and refers to the silicon vacancy.
$Distance$ refers to the distance between the silicon vacancy and the carbon antisite in \AA.
$Layer$ refers to the position of the carbon antisite with respect to the silicon vacancy, either inplane ($p$), below ($b$), and above ($a$) the plane of the silicon vacancy.
For example, the six closest configurations are denoted $h_{3.1}^{a}$,  $h_{3.1}^{p}$, $h_{3.1}^{b}$, $k_{3.1}^{a}$, $k_{3.1}^{p}$, and $k_{3.1}^{b}$.
These labels do not always correspond to a unique configuration, hence, in cases, where there are multiple non-equivalent atoms with the same distance and layer, these are separated with an index of I, II, or III; see Figure~\ref{fig:intro}b.

Carbon antisite positions along the c-axis are labeled with (layer $^{'}$ prime) to indicate the layer distance from the silicon vacancy plane.
From our compact nomenclature, the carbon antisite shares the same local environment ($h$ or $k$) with the silicon vacancy for even primes (such as $p$, $a^{''}$), whereas for odd primes (such as $a$, $a^{'''}$) the antisite is placed in different layer than the vacancy.
The ordering of $h$ and $k$ planes in 4H-SiC results in unique antisite positions for $h$ and $k$ vacancies.
The closest modified vacancy with C$_{\mathrm{3v}}$ symmetry for the $k$ silicon vacancy is a carbon antisite at 5 \AA\, whereas the nearest modified vacancy with C$_{\mathrm{3v}}$ symmetry for the $h$ configuration is at a distance of 10 \AA. 
As we explain later in Sec.~\ref{sec:ident}, the $k$ site modified vacancies with C$_{\mathrm{3v}}$ symmetry are the likely candidates for the unidentified defects in EPR experiments.

\section{Results}
\label{sec:res}

This section presents our high-accuracy computational results for the modified vacancy and experimental \ac{PL} measurements.
See Ref.~\ref{sec:method} for method details.
The theoretical results include formation energy, optical properties like \ac{ZPL} and \ac{TDM}, electronic structure, and \ac{ZFS}.
All the presented data is found in Table~\ref{tab:h} and Table~\ref{tab:k}.

\subsection{Theoretical Results}

\begin{figure}[h!]
   \includegraphics[width=0.52\columnwidth]{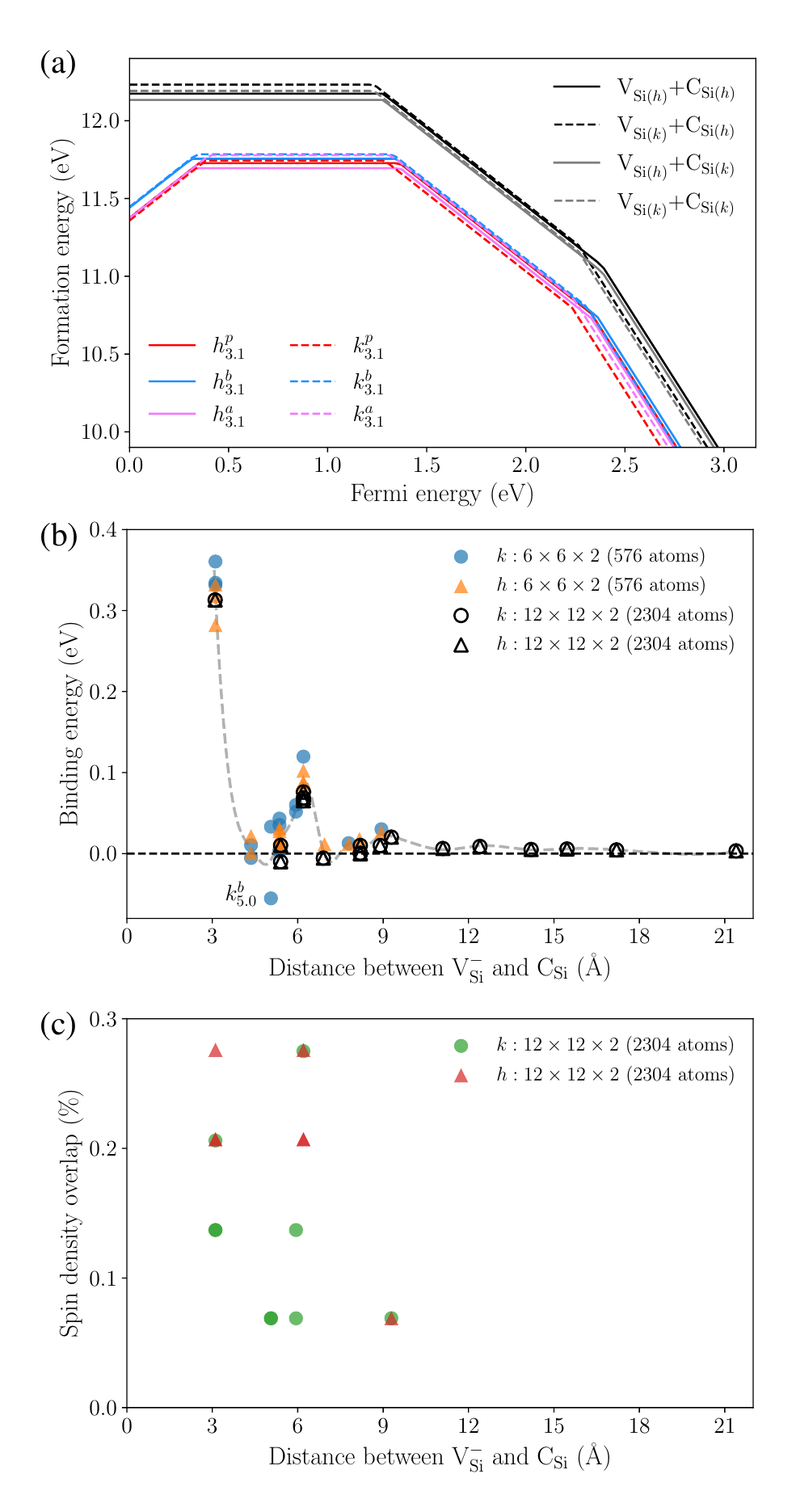}
	\caption{Energetics and the site dependent behavior of modified silicon vacancies. a) the formation energy (HSE results) for the closest modified vacancy configurations (colored lines) compared to the isolated silicon vacancy and carbon antisite (black and grey lines); b) the binding energy as a function of distance between the silicon vacancy and the carbon antisite. Notice the guideline to show the oscillating behavior of the binding energy. Here, the PBE results for the two different supercell sizes are shown. c) the spin density overlap (PBE results) for the carbon antisite on the isolated silicon vacancy defect states. Sites with large spin density overlap in (c) correspond to the local maxima in binding energy in (b).}
	\label{fig:form} 
\end{figure}

Figure~\ref{fig:form}a shows the formation energy of the six closest configurations of the modified silicon vacancies obtained by the HSE functional.
The formation energy trend for modified vacancies resembles silicon vacancy formation energy since the carbon antisite is neutral across all Fermi energies.
The only difference is that the modified vacancy has a stable positive charge state in contrast to the isolated silicon vacancy.
In the negative charge state, the binding energy for these defects is around 0.3 eV.
The positive binding energy and the fact that the modified vacancy is on the defect hull make it a stable defect.
Figure~\ref{fig:form}b shows the binding energy for all the considered negatively charged configurations in the 576 and 2304 atom supercells calculated with the PBE functional.
Two different supercells are used to ensure that the defect self-interaction is low for the configurations with large separation.
The perturbative effect of carbon antisite decreases as the antisite is placed farther away from the silicon vacancy, and the binding energy approaches zero beyond a separation of 10 Å.
However, there are peaks in the binding energy at periodic intervals of about 3 \AA\ (see the added guideline in Figure~\ref{fig:form}b).
The peaks appear when the carbon antisite is placed at a silicon site with a large spin density overlap, which is plotted in Figure~\ref{fig:form}c.
For an isolated silicon vacancy, the spin density is localized on carbon sites (Figure~\ref{fig:intro}b), and the placement of carbon antisite adjacent to spin density results in attractive interaction.
Two configurations also have negative binding energy, most notably $k^b_{5.0}$.

\begin{figure}[h!]
   \includegraphics[width=\columnwidth]{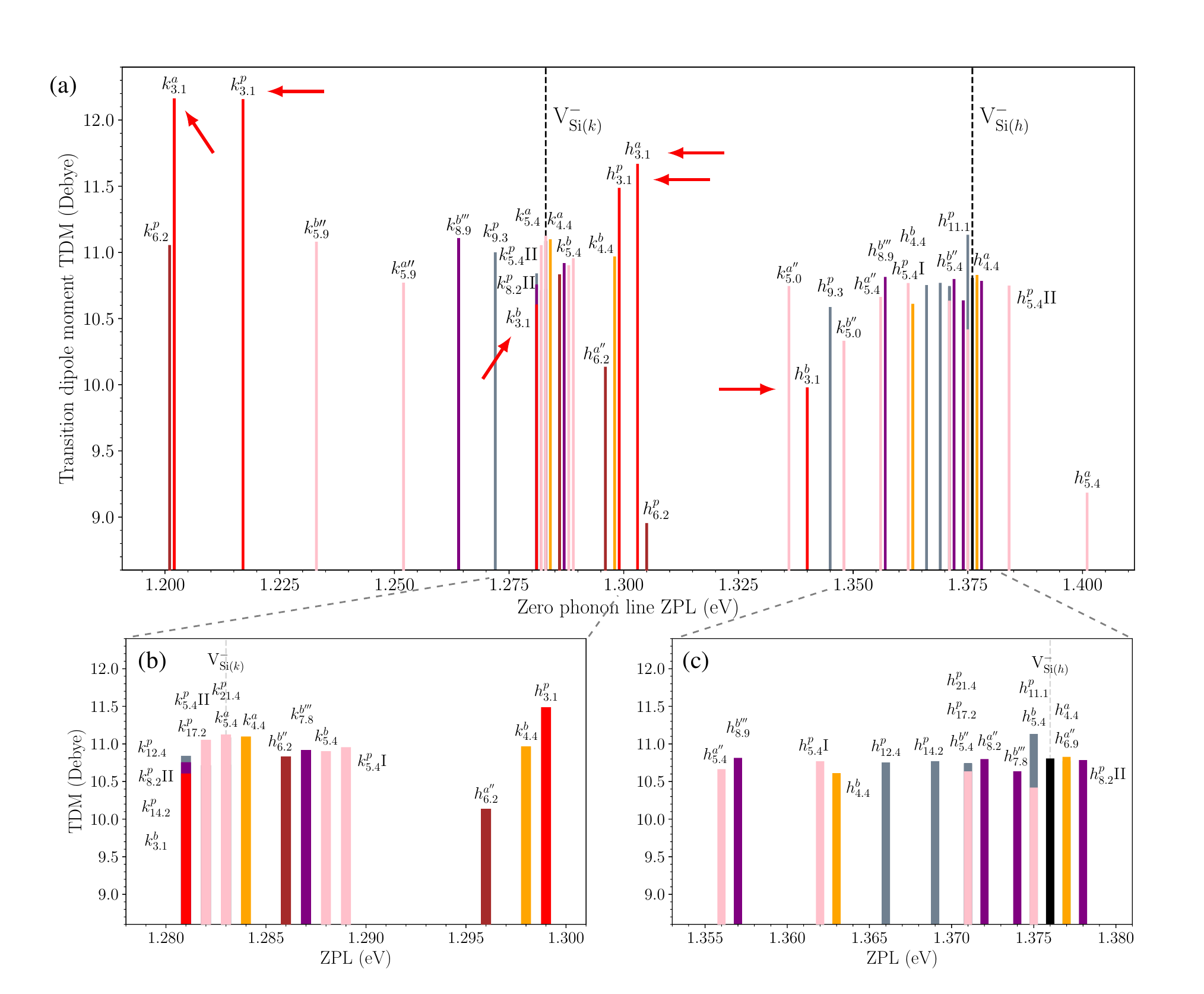}
	\caption{The optical data for the modified silicon vacancies. a) The ZPL (Both PBE and HSE results, the PBE results are shifted so that the isolated vacancies are align with the HSE results) and TDM for all the configurations of the modified silicon vacancies where the isolated silicon vacancies are marked with vertical dashed lines. The six closest configurations are marked with red arrows. b) and c) zoomed in versions close to the $k$ and $h$ isolated silicon vacancy, respectively.}
	\label{fig:zpl}
\end{figure}

Next, we discuss the optical properties of modified vacancies.
Figure~\ref{fig:zpl}a shows the ZPL and TDM for the modified silicon vacancies compared with the isolated silicon vacancies.
Here, one can see that the ZPL for the six closest modified silicon vacancies decreases by $\approx$75 meV, and the TDM increases by $\approx$1 Debye compared to the isolated counterparts.
As the antisite moves further away, the ZPL and TDM get closer to the isolated silicon vacancy, as shown in Figures~\ref{fig:zpl}b and c.

\begin{figure}[h!]
   \includegraphics[width=\columnwidth]{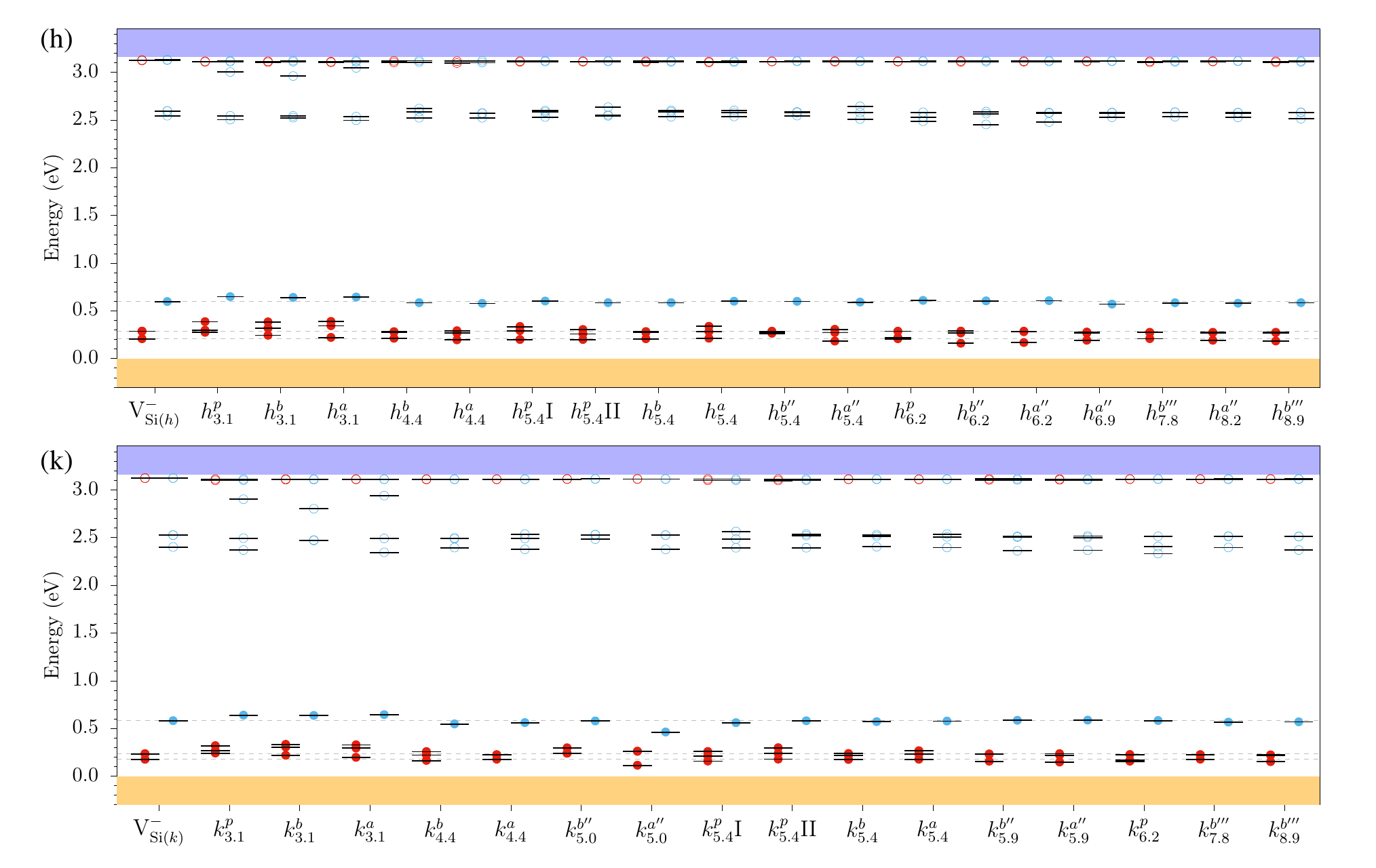}
	\caption{The Kohn-Sham eigenvalues (HSE results) for the modified vacancy for the h) and k) configurations compared with the isolated silicon defects. On the x-axis is the different configurations, where the subscript is the distance between the carbon antisite and the silicon vacancy. The dashed lines denote the eigenvalues for the occupied defect states of isolated silicon vacancy for both $h$ and $k$ configurations. The red and blue filled dots represent occupied states in each spin channel. Unfilled dots represents unoccupied states.}
	\label{fig:eigenvalue} 
\end{figure}

Figure~\ref{fig:eigenvalue} shows the eigenvalues for the modified vacancies.
As the carbon antisite is moved farther away from the silicon vacancy, the eigenvalues approach the isolated case.
The isolated silicon vacancy has three unoccupied states in one spin channel (marked in blue).
The lowest is a single degenerate $a_1$ state, and above it, there is a double degenerate $e$ state.
For the six closest configurations, the eigenvalues show the largest displacement with a prominent upward shift for the occupied states.
The unoccupied states have the largest splitting between the previous degenerate $e$ states.
Except for the below configurations ($h^{b}_{3.1}$ and $k^{b}_{3.1}$) that barely split the degenerate $e$ states but reorders the $a_1$ and $e$ states compared to the isolated case.
This outcome explains why the below configurations have a much lower ZPL change, see Figure~\ref{fig:zpl}a, compared to the above and planar configurations at the same distance.
At around 6 \AA, the eigenvalues are indistinguishable from the isolated case.
This agrees with similar trends for ZPL and binding energy, thus highlighting the extent of local site symmetry away from Si vacancy.
This eigenvalue shift is similar to the shift due to compressive strain of the isolated silicon vacancy~\cite{Udvarhelyi2020}.

\begin{figure}[h!]
   \includegraphics[width=\columnwidth]{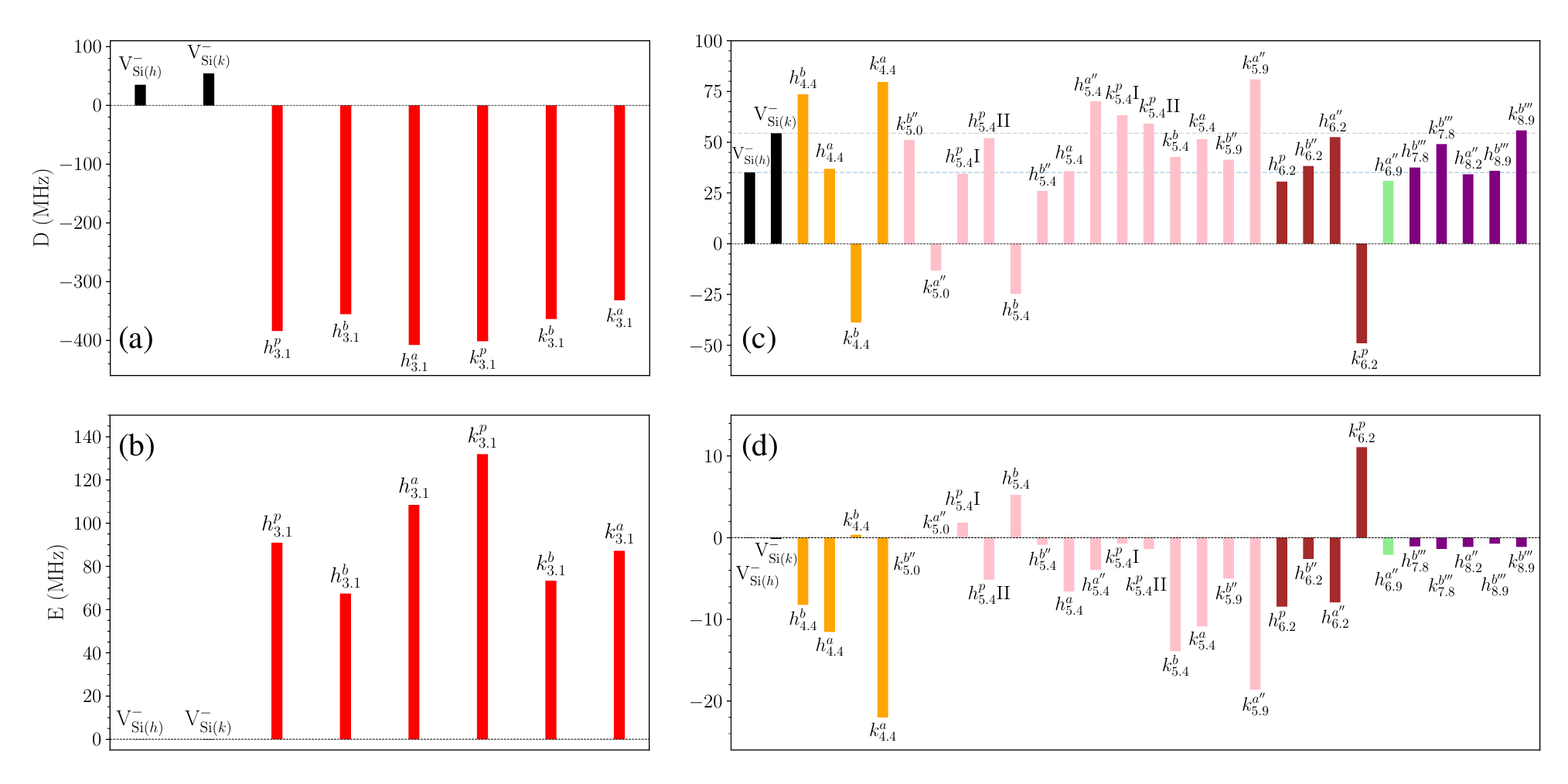}
	\caption{ZFS for the modified silicon vacancies (HSE results). a) the D value and b) the E value for the six closest configurations of the modified silicon vacancies compared with the isolated silicon vacancy; c) the D-tensor and d) the E-tensor for the remaining configurations. The data is grouped by distinct colors to easily identify the carbon antisite distance from the silicon vacancy.}
	\label{fig:zfs} 
\end{figure}

Figure~\ref{fig:zfs} shows the ZFS for the modified vacancy compared to the isolated silicon vacancy.
The six closest configurations have much larger E and D values than the isolated case.
However, as the distance between the defects increases, the values approach the isolated silicon values due to the exponential decay of the overlapping spin densities.
This change corresponds well with the difference in the eigenvalues as the distance increases.

\renewcommand{\arraystretch}{0.94}

\begin{table}[h!]
\caption{The data for all the modified vacancy $h$ configurations. Each configuration is denoted with the symmetry and distance between the silicon vacancy and carbon antisite as well as binding energy. The magneto-optical properties include ZPL (for both 576 atom super cell with the HSE functional and the 2304 super cell with the PBE functional), TDM, D and E values.}
\begin{ruledtabular}
\begin{tabular} {c|cr|r|rr|rrr|rr|r}
\multirow{2}{*}{\rotatebox[origin=c]{90}{Defect}} & \multirow{2}{*}{\rotatebox[origin=c]{90}{Symmetry}}  & Distance & Binding & ZPL HSE & ZPL PBE & \multicolumn{3}{c|}{TDM (Debye)} & D (MHz) & E (MHz)   & Spin density  \\
& & (\AA) & Energy (eV) & 576 (eV) & 2304 (eV) & perp & para & tot &  &  & overlap (\%) \\
\hline
$h^{p}_{3.1}$               &   C$_{1h}$   & 3.1              & 0.332 & 1.299     & 1.168  &    6.0  &  9.8  & 11.5         &  -384.09   &    91.00    & 0.207 \\
$h^{b}_{3.1}$               &   C$_{1h}$   & 3.1              & 0.282 & 1.340     &   &        10.0  & 0.3   & 10.0         &  -355.47   &    67.46    & 0.276 \\
$h^{a}_{3.1}$               &   C$_{1h}$   & 3.1              & 0.318 & 1.303     &   &         5.1  & 10.4  & 11.7         &  -407.71   &    108.42   & 0.207 \\
$h^{a}_{4.4}$               &   C$_{1h}$   & 4.4              & 0.001   & 1.377     &   &         1.1  & 10.8  & 10.8         &   36.84    &   -11.55    &  \\
$h^{b}_{4.4}$               &   C$_{1h}$   & 4.4              & 0.021 & 1.363     &   &         4.4  & 9.6   & 10.6         &   73.68    &   -8.24     &  \\
$h^{p}_{5.4}\textrm{I}$     &   C$_{1h}$   & 5.4              & 0.010 & 1.362     & 1.277  &    0.2  & 10.7  & 10.7         &   34.44    &    1.84     &  \\
$h^{p}_{5.4}\textrm{II}$    &   C$_{1h}$   & 5.4              & 0.031 & 1.384     &   &         0.6  & 10.7  & 10.7         &   51.99    &   -5.13     &  \\
$h^{b}_{5.4}$               &   C$_{1h}$   & 5.4              & 0.027 & 1.375     &   &         5.2  &  9.0  & 10.4         &  -24.64    &    5.23     &  \\
$h^{a}_{5.4}$               &   C$_{1h}$   & 5.4              & 0.015 & 1.401     &   &         9.2  &  0.0    &  9.2         &   35.77    &   -6.56     &  \\
$h^{b''}_{5.4}$             &   C$_{1h}$   & 5.4              & 0.009 & 1.371     &   &         0.6  & 10.6  & 10.6         &   26.01    &   -0.87     &  \\
$h^{a''}_{5.4}$             &   C$_{1h}$   & 5.4              & 0.019 & 1.356     &   &         0.9  & 10.6  & 10.6         &   70.23    &   -3.95     &  \\
$h^{p}_{6.2}$               &   C$_{1h}$   & 6.2              & 0.102 & 1.305     & 1.162  &    8.9  &  0.9  &  8.9         &   30.61    &   -8.44     & 0.276 \\
$h^{b''}_{6.2}$             &   C$_{1h}$   & 6.2              & 0.086 & 1.286     & 1.160  &    1.0  & 10.8  & 10.8         &   38.26    &   -2.59     & 0.207 \\
$h^{a''}_{6.2}$             &   C$_{1h}$   & 6.2              & 0.088 & 1.296     & 1.166  &    5.0  &  8.8  & 10.1         &   52.50    &   -7.91     & 0.207 \\
$h^{a''}_{6.9}$             &   C$_{1h}$   & 6.9              & 0.011 & 1.377     & 1.243  &    5.2  &  9.4  & 10.7         &   30.93    &   -2.09     &  \\
$h^{b'''}_{7.8}$            &   C$_{1h}$   & 7.8              & 0.012 & 1.374     &        &    3.6  & 10.0  & 10.6         &   37.53    &   -1.06     &  \\
$h^{a''}_{8.2}$             &   C$_{1h}$   & 8.2              & 0.018 & 1.372     & 1.237  &    5.3  &  9.3  & 10.7         &   34.20    &   -1.12     &  \\
$h^{b'''}_{8.9}$            &   C$_{1h}$   & 8.9              & 0.027 & 1.357     & 1.226  &    4.3  &  9.8  & 10.7         &   35.96    &   -0.73     &  \\
$h^{p}_{8.2}\textrm{II}$    &   C$_{1h}$   & 8.2              & 0.001   &           & 1.247  &    2.8  & 10.3  & 10.7         &            &             &  \\
$h^{p}_{9.3}$               &   C$_{1h}$   & 9.3              & 0.020 &           & 1.214  &   10.5  &  0.8  & 10.6         &            &             & 0.069 \\
$h^{p}_{11.1}$              &   C$_{1h}$   & 11.1             & 0.006 &           & 1.244  &    9.0  &  6.5  & 11.1         &            &             &  \\
$h^{p}_{12.4}$              &   C$_{1h}$   & 12.4             & 0.009 &           & 1.214  &    3.5  & 10.2  & 10.7         &            &             &  \\
$h^{p}_{14.2}$              &   C$_{1h}$   & 14.2             & 0.005 &           & 1.238  &    5.4  &  9.3  & 10.8         &            &             &  \\
$h^{p}_{17.2}$              &   C$_{1h}$   & 17.2             & 0.004 &           & 1.240  &    5.4  &  9.3  & 10.7         &            &             &  \\
$h^{p}_{21.4}$              &   C$_{1h}$   & 21.4             & 0.003 &           & 1.240  &    5.3  &  9.4  & 10.7         &            &             &  \\
V$^{-}_{\mathrm {Si}(h)}$   &   C$_{3v}$   &                  &       & 1.376     & 1.245  &    0.5  & 10.8  & 10.8         &   35.10    &   -0.04   & \\
\end{tabular}
\end{ruledtabular}
\label{tab:h}
\end{table}

\begin{table}[h!]
\caption{The data for all the modified vacancy $k$ configurations. Each configuration is denoted with the symmetry and distance between the silicon vacancy and carbon antisite as well as binding energy. The magneto-optical properties include ZPL (for both 576 atom super cell with the HSE functional and the 2304 super cell with the PBE functional), TDM, D and E values.}
\begin{ruledtabular}
\begin{tabular} {c|cr|r|rr|rrr|rr|r}
\multirow{2}{*}{\rotatebox[origin=c]{90}{Defect}} & \multirow{2}{*}{\rotatebox[origin=c]{90}{Symmetry}}  & Distance & Binding & ZPL HSE & ZPL PBE & \multicolumn{3}{c|}{TDM (Debye)} & D (MHz) & E (MHz)   & Spin density  \\
& & (\AA) & Energy (eV) & 576 (eV) & 2304 (eV) & perp & para & tot &  &  & overlap (\%) \\
\hline
$k^{p}_{3.1}$               &   C$_{1h}$   & 3.1              & 0.360 & 1.217     & 1.110  &    5.9  & 10.6  & 12.1         &  -401.59   &    131.92   & 0.206 \\
$k^{b}_{3.1}$               &   C$_{1h}$   & 3.1              & 0.331 & 1.281     &   &        10.6  &  0.7  & 10.6         &  -363.70   &    73.34    & 0.137 \\
$k^{a}_{3.1}$               &   C$_{1h}$   & 3.1              & 0.334 & 1.202     &   &         4.2  & 11.4  & 12.1         &  -331.93   &    87.21    & 0.137 \\
$k^{a}_{4.4}$               &   C$_{1h}$   & 4.4              & -0.005 & 1.284    &   &         0.0    & 11.0  & 11.0         &   79.68    &   -22.01    &  \\
$k^{b}_{4.4}$               &   C$_{1h}$   & 4.4              & 0.010 & 1.298     &   &         1.1  & 10.9  & 10.9         &  -38.87    &    0.35     &  \\
$k^{a''}_{5.0}$             &   C$_{3v}$   & 5.0              & 0.033 & 1.336     &   &         0.0    & 10.7  & 10.7         &  -13.34    &    0.05     & 0.069 \\
$k^{b''}_{5.0}$             &   C$_{3v}$   & 5.0              & -0.055 & 1.348    &   &         5.8  &  8.5  & 10.3         &   51.06    &   -0.11     & 0.069 \\
$k^{p}_{5.4}\textrm{I}$     &   C$_{1h}$   & 5.4              & 0.043 & 1.289     & 1.177  &    0.7  & 10.9  & 10.9         &   63.28    &   -0.70     &  \\
$k^{p}_{5.4}\textrm{II}$    &   C$_{1h}$   & 5.4              & 0.004 & 1.282     &   &         0.7  & 11.0  & 11.0         &   59.12    &   -1.39     &  \\
$k^{b}_{5.4}$               &   C$_{1h}$   & 5.4              & 0.035 & 1.288     &   &         4.7  &  9.8  & 10.9         &   42.67    &   -13.87    &  \\
$k^{a}_{5.4}$               &   C$_{1h}$   & 5.4              & 0.002 & 1.283     &   &         0.5  & 11.1  & 11.1         &   51.49    &   -10.85    &  \\
$k^{b''}_{5.9}$             &   C$_{1h}$   & 5.9              & 0.051 & 1.233     &   &         1.3  & 11.0  & 11.1         &   41.39    &   -5.00     & 0.137 \\
$k^{a''}_{5.9}$             &   C$_{1h}$   & 5.9              & 0.060 & 1.252     &   &         3.0  & 10.3  & 10.7         &   80.99    &   -18.60    & 0.069 \\
$k^{p}_{6.2}$               &   C$_{1h}$   & 6.2              & 0.119 & 1.201     & 1.096  &    7.6  &  8.0  & 11.0         &  -48.98    &    11.08    & 0.275 \\
$k^{b'''}_{7.8}$            &   C$_{1h}$   & 7.8              & 0.013 & 1.287     &   &         0.1  & 10.9  & 10.9         &   49.10    &   -1.36     &  \\
$k^{b'''}_{8.9}$            &   C$_{1h}$   & 8.9              & 0.030 & 1.264     &   &         0.6  & 11.1  & 11.1         &   55.78    &   -1.12     &  \\
$k^{p}_{8.2}\textrm{II}$    &   C$_{1h}$   & 8.2              & 0.014 &           & 1.173  &    1.3  & 10.7  & 10.8         &            &             &  \\
$k^{p}_{9.3}$               &   C$_{1h}$   & 9.3              & 0.023 &           & 1.173  &    5.5  &  9.5  & 11.0         &            &             & 0.069 \\
$k^{p}_{11.1}$              &   C$_{1h}$   & 11.1             & 0.003 &           & 1.174  &   10.9  &  0.2  & 10.9         &            &             &  \\
$k^{p}_{12.4}$              &   C$_{1h}$   & 12.4             & 0.010 &           & 1.173  &    3.5  & 10.2  & 10.8         &            &             &  \\
$k^{p}_{14.2}$              &   C$_{1h}$   & 14.2             & 0.004 &           & 1.174  &    1.5  & 10.6  & 10.7         &            &             &  \\
$k^{p}_{17.2}$              &   C$_{1h}$   & 17.2             & 0.004 &           & 1.174  &    0.3  & 10.7  & 10.7         &            &             &  \\
$k^{p}_{21.4}$              &   C$_{1h}$   & 21.4             & 0.005 &           & 1.175  &    1.3  & 10.6  & 10.7         &            &             &  \\
V$^{-}_{\mathrm {Si}(k)}$   &   C$_{3v}$   &                  &       & 1.283     & 1.175  &   0.4   & 11.0  & 11.0         &   54.38    &   -0.16     &  \\
\end{tabular}
\end{ruledtabular}
\label{tab:k}
\end{table}

\subsection{Experimental \ac{PL} Measurements}

\begin{figure}[h!]
   \includegraphics[width=0.4\columnwidth]{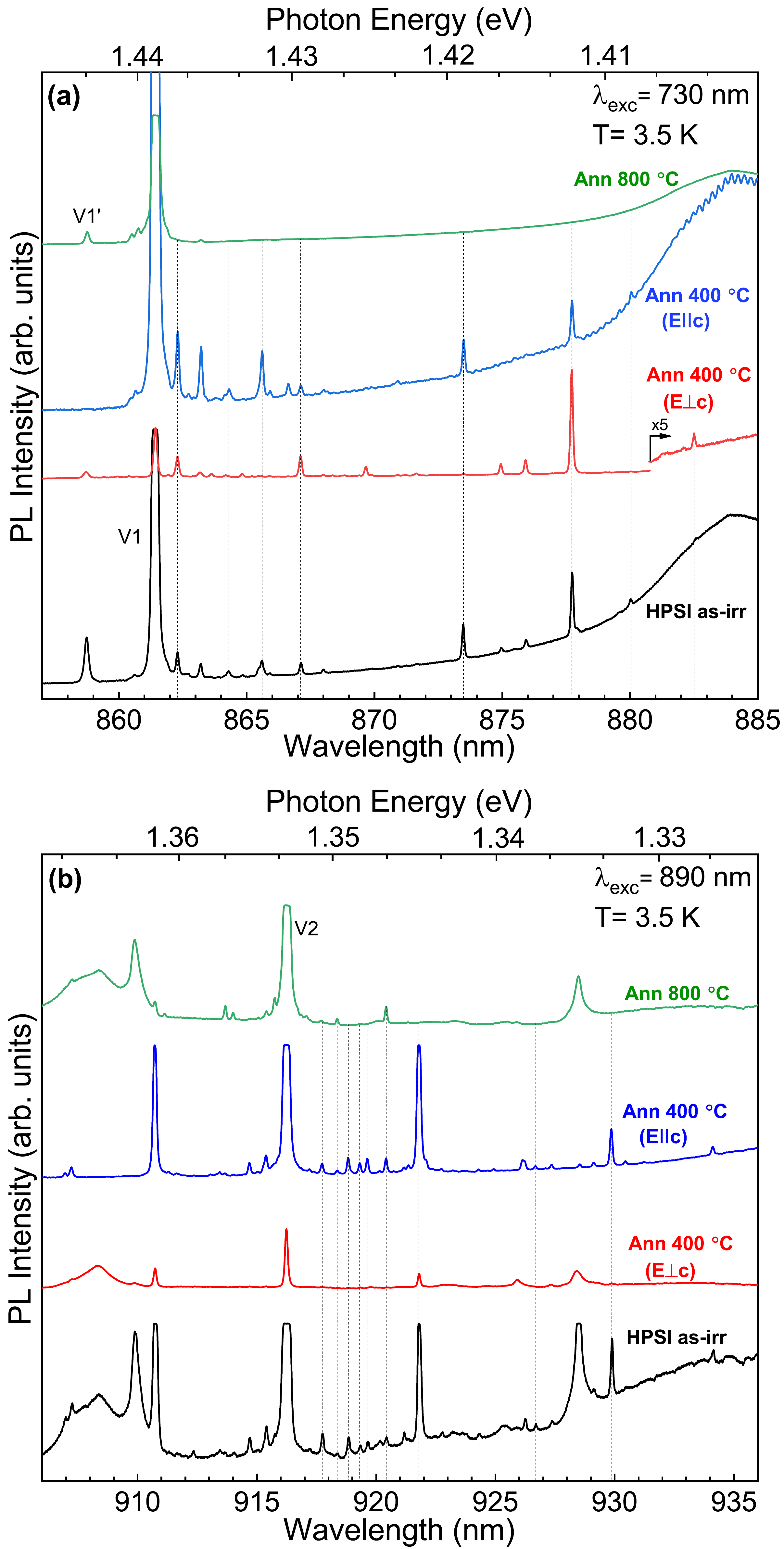}
	\caption{PL for the high-purity semi-insulating 4H-SiC samples. a) show the range of 860-885 nm containing the V1 and V1' signals and b) show the range of 910-935 nm containing the V2 signal. The figures show there are several small additional lines next to the V1 and V2 lines. These additional lines appear in as-irradiated samples and only some faint signals remain after annealing at 800 $^{\circ}$C. The polarization is also shown for the 400 $^{\circ}$C annealed sample.}
	\label{fig:exp}
\end{figure}

In this section, we present results from three high-purity semi-insulating (HPSI) 4H-SiC samples, which all exhibit strong PL from silicon vacancy.
See Section.~\ref{sec:ex_method} for experimental details.
The PL spectra are displayed in Figure~\ref{fig:exp} which shows several additional lines in a $\approx$20 nm range in the vicinity of each the isolated the $h$ and $k$ silicon vacancies.
We have counted up to 63 additional lines which are listed in Table~\ref{tab:exp_h} and Table~\ref{tab:exp_k} with the polarization given for the most prominent lines.

\renewcommand{\arraystretch}{0.73}

\begin{table}[h!]
\caption{The \ac{PL} data for lines around the $h$ isolated silicon vacancies in 4H-SiC. The polarization is normalized depending on the photon count and only shown for bright lines. Only mixed polarization with more than 5\% in the minority component are shown. See Sec.~\ref{sec:ident} for more information about the assigned configurations in the comment column.}
\begin{ruledtabular}
\begin{tabular} {cc|rr|c}
\multicolumn{2}{c|}{Line}  & \multicolumn{2}{c|}{Polarization}  & Comment \\
(nm) & (eV) & perp (\%) & para (\%)  \\
\hline
858.7	&	1.444	&	100	&	0	&	V1'	\\
860.7	&	1.441	&		&		&		\\
861.4	&	1.440	&	0	&	100	&	V1	\\
862.2	&	1.438	&	24	&	76	&	$h^{a''}_{6.9}$	\\
862.7	&	1.437	&	0	&	100	&		\\
862.9	&	1.437	&	0	&	100	&		\\
863.2	&	1.437	&	10	&	90	&	$h^{b'''}_{7.8}$	\\
863.6	&	1.436	&	100	&	0	&	$h^{a}_{5.4}$	\\
864.1	&	1.435	&	48	&	52	&	$k^{p}_{6.2}$	\\
864.3	&	1.435	&	0	&	100	& $h^{b''}_{5.4}$ \\
864.8	&	1.434	&	100	&	0	& $h^{p}_{9.3}$	 \\
865.6	&	1.433	&	0	&	100	&	\\
865.9	&	1.432	&	0	&	100	&	\\
866.6	&	1.431	&	0	&	100	&	\\
867.1	&	1.430	&	77	&	23	&		\\
867.9	&	1.429	&		&		&		\\
868.3	&	1.428	&		&		&		\\
868.8	&	1.427	&		&		&		\\
869.1	&	1.427	&		&		&		\\
869.4	&	1.426	&		&		&		\\
869.6	&	1.426	&	0	&	100	&	$h^{p}_{5.4}\textrm{I}$	\\
869.8	&	1.426	&		&		&		\\
870.9	&	1.424	&		&		&		\\
871.2	&	1.423	&		&		&		\\
871.6	&	1.423	&		&		&		\\
872.4	&	1.421	&		&		&		\\
873.4	&	1.420	&	0	&	100	&	\\
874.9	&	1.417	&	100	&	0	&		\\
875.2	&	1.417	&		&		&		\\
875.9	&	1.416	&	100	&	0	&	\\
877.7	&	1.413	&	71	&	29	&		\\
880.0	&	1.409	&	0	&	100	&	$k^{a''}_{5.0}$	\\
882.5	&	1.405	&	100	&	0	&	$h^{b}_{3.1}$	\\
\end{tabular}
\end{ruledtabular}
\label{tab:exp_h}
\end{table}

\begin{table}[h!]
\caption{The \ac{PL} data for lines around the $k$ isolated silicon vacancies in 4H-SiC. The polarization is normalized depending on the photon count and only shown for bright lines. Only mixed polarization with more than 5\% in the minority component are shown. See Sec.~\ref{sec:ident} for more information about the assigned configurations in the comment column.}
\begin{ruledtabular}
\begin{tabular} {cc|rr|c}
\multicolumn{2}{c|}{Line}  & \multicolumn{2}{c|}{Polarization} & Comment \\
(nm) & (eV) & perp (\%) & para (\%)  \\
\hline
910.7	&	1.362	&	0	&	100	&	\\
911.2	&	1.361	&	0	&	100	&		\\
913.0	&	1.358	&	0	&	100	&	\\
913.4	&	1.358	&	0	&	100	& $k^{b'''}_{7.8}$	\\
914.0	&	1.357	&		&		&		\\
914.6	&	1.356	&	0	&	100	&		\\
915.3	&	1.355	&	0	&	100	& $k^{a}_{4.4}$	\\
916.3	&	1.353	&	0	&	100	&	V2	\\
917.2	&	1.352	&		&		&		\\
917.4	&	1.352	&		&		&		\\
917.7	&	1.351	&	0	&	100	&		\\
918.3	&	1.350	&	0	&	100	&		\\
918.8	&	1.350	&	0	&	100	&		\\
919.3	&	1.349	&	0	&	100	&		\\
919.6	&	1.348	&	0	&	100	&		\\
920.1	&	1.348	&		&		&		\\
920.4	&	1.347	&		&		&		\\
920.7	&	1.347	&	0	&	100	&		\\
921.2	&	1.346	&		&		&		\\
921.3	&	1.346	&	0	&	100	&		\\
921.7	&	1.346	&	0	&	100	&		\\ 
922.7	&	1.344	&		&		&		\\
923.2	&	1.343	&		&		&		\\
924.3	&	1.342	&		&		&		\\
924.9	&	1.341	&		&		&		\\
926.1	&	1.339	&	0	&	100	&		\\
926.7	&	1.338	&		&		&		\\
927.1	&	1.338	&		&		&		\\
927.3	&	1.337	&		&		&		\\
928.5	&	1.335	&		&		&		\\
929.1	&	1.335	&		&		&		\\
929.8	&	1.334	&	0	&	100	&	$k^{b'''}_{8.9}$	\\
930.4	&	1.333	&		&		&		\\
931.2	&	1.332	&		&		&		\\
\end{tabular}
\end{ruledtabular}
\label{tab:exp_k}
\end{table}

\section{Discussion}
\label{sec:dis}

\subsection{General Trends}

The magneto-optical properties of the modified vacancy depend on the position of the carbon antisite.
As the carbon antisite moves farther away from the silicon vacancy, one would expect the ZPL to approach the isolated silicon vacancy.
Surprisingly, at around 6 \AA, the ZPL shift is as significant as for the closest configurations, see Figure~\ref{fig:zpl}a.
This trend is present for both supercell sizes and functional used, and is explained by the overlap of the silicon vacancy spin densities with the position of the antisites.
At 3.1 \AA\ and 6.2 \AA\ away from the vacancy, the overlap is large.
When carbon atoms are replaced on these silicon sites, they significantly affect the total energy, Kohn-Sham eigenvalues, \ac{ZFS}, and \ac{ZPL}.
For the ZFS, the closest configurations have an order of magnitude larger D values compared to the isolated silicon vacancy.
However, the low symmetry makes it hard to detect.

As the carbon antisite moves farther away, the defect orbitals around the silicon vacancy are modified less, and the values resemble the isolated silicon vacancy.
We argue that even if the global symmetry is low (which makes it difficult to measure EPR signal~\cite{Adam_mod_Si}), the local site symmetry around the silicon vacancy still remains high (C$_{\mathrm{3v}}$) for certain configurations.
One can find an example of this in Table~\ref{tab:h} and Table~\ref{tab:k}, the 3rd nearest neighbor site $k^{b}_{4.4}$ has C$_{\mathrm{1h}}$ symmetry but a low E value indicates that the electronic structure is close to C$_{\mathrm{3v}}$ symmetry, see Figure~\ref{fig:eigenvalue}.
Thus, the presence of electronically benign defects such as antisites near spin defects should be studied with regards to the orientation of spin density.

Of the tested modified vacancy configurations, two have C$_{\mathrm{3v}}$ symmetry within 10 \AA.
The $k^{a''}_{5.0}$ configuration has the smallest D value ($-13.34$, see Table~\ref{tab:k} or Figure~\ref{fig:zfs}c) of all the tested configurations and could be the defect responsible for the $R_1$ EPR signal~\cite{Son_2019_Ligand}.
The same configuration (named $\mathrm{V_{Si}^-}$-$\mathrm{C_{Si}}(1)$-$k$ in Ref.~\onlinecite{Adam_mod_Si}) also has the lowest D value.
We draw the same conclusion as the authors in Ref.~\onlinecite{Adam_mod_Si}, the $k^{a''}_{5.0}$ configuration is responsible for $R_1$ EPR signal.
With the additional configurations considered in this paper, it is clear that no other configuration has a smaller ZFS value than $k^{a''}_{5.0}$.
The other modified vacancy with C$_{\mathrm{3v}}$ symmetry, $k^{b''}_{5.0}$, was not considered in Ref.~\onlinecite{Adam_mod_Si}.
It has a large ZFS which could be related to $T_{V1b}$.
However, it has negative binding energy.
Several other candidates could fit this measurement; like $k^{a}_{5.4}$, $h^{p}_{5.4}$II, and $h^{a''}_{6.2}$II for example; but they have larger E values.
Further experiments are needed before any conclusion can be drawn.
The C$_{\mathrm{3v}}$ candidates for $h$ vacancy are $\approx$10 \AA\ away and are unlikely to have distinct \ac{EPR} signals.

The modified vacancies are created under irradiation conditions, and it is crucial to understand the behavior of these defect complexes at high temperatures.
Importantly, at what temperature would the modified vacancies anneal out?
Even if their binding energies are small, the defects must be mobile to separate.
The energy barriers for the different single defects depend on the charge state of the defect.
But the barrier ranges are around 1-2 eV for the interstitials, 2-5 eV for the carbon vacancy, 3-4 eV for the silicon vacancy, and $\approx$10 eV for both antisites in 4H-SiC~\cite{4Hdiffusion}.
The large energy barrier makes the antisites stable regardless of the annealing temperature.
The silicon vacancies would be mobile at high temperatures and either encounter a silicon antisite and transform to a carbon vacancy or bind to a carbon antisite creating the modified silicon vacancies discussed in this paper.
However, the carbon vacancies become mobile before the silicon vacancies due to the lower energy barrier.
They could interact with either the silicon vacancies to form divacancies or the carbon antisites to form silicon vacancies.
Both these cases would affect the modified vacancies.
The carbon vacancies become mobile at around 800 K (found with transition state theory~\cite{PhysRevB.97.214104,VINEYARD1957121}).
In Figure~\ref{fig:exp}, the additional lines disappear between 400 and 800 $^\circ$C annealing.
Hence, we put forward that the annealing behavior of the additional lines is due to the mobile carbon vacancies that interact with the modified vacancies.
The carbon vacancy either interacts with the silicon vacancy which transforms the modified vacancies into modified divacancies~\cite{mod_diV} or interacts with the carbon antisite, turning the modified vacancy into an isolated silicon vacancy.

Are there other defects that can explain the experimental measurements?
If these signals correspond to an unknown point defect containing a silicon vacancy, the modified vacancy with a carbon antisite is the only valid candidate considering double defects at close distances.
The number of spin-3/2 point defect clusters with silicon vacancies in the database of 8355 defects is low.
This extensive search shows that the modified vacancy is the most likely defect.
It has stable stoichiometry.
At larger distances, one could argue that any perturbing defect could give rise to the results presented in this paper.
However, there are a few factors that limit such scenarios.
Let us first consider an antisite that modifies the silicon vacancies.
Even though the carbon and silicon antisite have similar formation energies (carbon is slightly lower), the clusters formed by a silicon vacancy and silicon antisite have the same stoichiometry as a carbon vacancy and most likely will anneal to that defect.
The modified vacancy has a stoichiometry that cannot transform into any other single or double defect.
Similar reasoning can be applied for the interstitials: a carbon interstitial and silicon vacancy would transform into a carbon antisite, and a silicon interstitial and silicon vacancy (a Frenkel pair) would cancel each other out.
In the case of silicon vacancy modified with a carbon vacancy, it would lead to the formation of a divacancy.
A cluster with two silicon vacancies does not bind together~\cite{defect_hull}.
This only leaves the modified vacancy with a carbon antisite as a probable candidate of the considered defects.

\subsection{Identification}
\label{sec:ident}

Here, we compare the theoretical and experimental data on magneto-optical properties of the investigated defect complexes to look for matches.
Due to their larger \ac{TDM}, the modified vacancies would show stronger PL signals than the isolated silicon vacancies if the concentrations were the same.
However, since the modified vacancy is a double defect consisting of a carbon antisite and silicon vacancy, the concentration of modified vacancies will expectedly be lower than isolated silicon vacancy in irradiated and annealed samples.
Hence, we expect the modified vacancies to appear with lower intensity than the silicon vacancies, which is the case for the new lines in Figure~\ref{fig:exp}.
The concentration differences does not affect the identification.

The identification is mainly based on the \ac{PL} values and supported with \ac{EPR} values when available.
For the PL lines without ZFS, the assignment is done by comparing the ZPL values and polarization.
The theoretical HSE ZPL values are shifted to align the isolated silicon vacancies and remove systematic errors.
The relative errors between configurations are in the meV range~\cite{methodology_paper}.
Then the least squared distance (lsd) between theoretical ZPL (eV), para (fraction), perp (fraction) and the experimental values is found with the following equation:
\begin{equation}
\mathrm{lsd = \big(ZPL_{exp} - ZPL^{shifted}_{theory}  \big)^2 +  \big(para_{exp} - para_{theory}  \big)^2 +  \big(perp_{exp} - perp_{theory}  \big)^2}
\label{eq:lsd}
\end{equation}
When one theoretical configuration matches with multiple experimental values, the smallest lsd value is used to assign the line.
The assigned configurations are summarized in Table~\ref{tab:id}, and also marked in Table~\ref{tab:exp_h} and Table~\ref{tab:exp_k}.

Many of the predicted ZPLs from the modified vacancies fall close to the isolated value.
This makes it difficult to exclude other effects like strain~\cite{Udvarhelyi2020}, L-lines~\cite{L-lines} or surface effects.
Hence, we first focus on the lines farthest away from the silicon vacancy, at least 15 meV from the isolated line.
The primary candidates for the $R_1$ EPR signal, $k^{a''}_{5.0}$ has ZPL around 40 meV from the isolated $h$ silicon vacancy with a parallel polarization, see Table~\ref{tab:k} and Figure~\ref{fig:zfs}a.
This is in excellent agreement with the 880.0 nm line observed in PL experiments with an identical polarization (Figure~\ref{fig:exp} a).
Furthermore, this line is also seen in the same sample used to measure the $R_1$ signal~\cite{Son_2019_Ligand}.
Hence, the $k^{a''}_{5.0}$ configuration is identified as the source of the $R_1$ signal and 880.0 nm line.

Next, we consider the five lines with perpendicular polarization (863.6, 864.8, 874.9, 875.9, and 882.5 nm) in Figure~\ref{fig:exp}a.
Table~\ref{tab:id} shows that the $h^{b}_{3.1}$ configuration is the ideal candidate for the 882.5 nm line (also suggested for 874.9 and 875.9 nm).
This complete change of polarization is due to the reordering of the $a_1$ and $e$ states compared to the isolated case, mentioned in Sec.~\ref{sec:res}.
Furthermore, the 863.6 nm line is best matched to the $h^{a}_{5.4}$ configuration and 864.8 nm to $h^{p}_{9.3}$.
The latter identification uses theoretical PBE ZPL data from the 2304 atoms to avoid nonphysical defect-defect interactions of the smaller supercell.
Both the $h^{b}_{3.1}$ and $h^{p}_{9.3}$ configurations lie on local maxima in the binding energy, see Figure~\ref{fig:form}b.

There are also several additional lines around the isolated $k$ silicon vacancy, see Figure~\ref{fig:exp}b.
In contrast to the $h$ configurations, the PL signals are observed on both sides of the isolated vacancy line V2.
Again, focusing at least 15 meV away from the isolated line, the 929.8 nm line is best described by the $k^{b'''}_{8.9}$ configuration (also suggested for 926.1 nm).
Closer to the isolated vacancy, there are two bright peaks on either side: 910.7 and 921.7 nm, which both are missing good candidates.
Since the peaks disappear upon annealing at high temperatures, they may correspond to structural defects.
Coincidentally, both peaks are almost equally spaced from the V2 line by 8.1 meV.
Furthermore, the 915.3 nm line is best described by the $k^{a}_{4.4}$ configuration.
The $k^{b'''}_{7.8}$ configuration could be any of the 910.7, 911.2, 913.0, 913.4, or 914.6 nm lines since they are close and have the same polarization.
It is tentatively assigned to the 913.4 nm line, which has the smallest lsd.

Within the 15 meV range next to the isolated $h$ silicon vacancy, we observe a set of lines in the same region as L-lines~\cite{L-lines}, see Figure~\ref{fig:exp}a.
However, there is no one-to-one correspondence due to disagreement in the observed intensities and energy spacing from earlier study~\cite{L-lines}.
The L-lines are tentatively suggested to be vibronic replicas of the silicon vacancy~\cite{L-lines}.
However, the L-lines disappear upon annealing at 800 K, whereas the silicon vacancy remains~\cite{L-lines} which contradicts this hypothesis.
Alternatively, these lines could originate from strain, surface effects, or another defect.
Several possible modified vacancy candidates from Table~\ref{tab:h} match the experimental spectra.
By comparing the HSE results of the modified vacancies (Table~\ref{tab:h}) with experiment (Table~\ref{tab:exp_h}), the best matches for both ZPL energy and polarization are:
862.2 nm is $h^{a''}_{6.9}$ and 863.2 nm is $h^{b'''}_{7.8}$.
The 864.1 nm is matched to $k^{p}_{6.2}$ with excellent polarization but the ZPL match is terrible, which is reflected in the high lsd.
This result is a product of the automatic matching.
$h^{p}_{5.4}\textrm{I}$ is either 866.6 nm, 869.6 nm (assigned), or 873.4 nm, and $h^{b''}_{5.4}$ could be either of 864.3 nm (assigned), 865.6 nm, or 865.9 nm.
The 862.7 and 862.9 nm are too similar to the isolated silicon vacancy for a definite assignment.
Note that we did not find any good matches between theory and experiment for the 867.1 and 877.7 nm lines with mixed polarization.
Further studies of these lines are needed.

\begin{table}[h!]
\caption{Summary of the identified modified vacancy configurations. In this table, the theoretical HSE ZPL results have been shifted to align with the experimental isolated silicon vacancy results for easier comparison. This means a shift of 64 meV for the $h$ configurations and 70 meV for the $k$ configurations. Unshifted values are presented in Table~\ref{tab:exp_h} and Table~\ref{tab:exp_k}. The assigned configurations are marked in bold. When multiple configurations match an experimental line, the configuration with the smallest least squared is assigned.}
\begin{ruledtabular}
\begin{tabular} {cc|rr|cc|rr|c|c}
\multicolumn{4}{c|}{Experimental \ac{PL} Data}  & \multicolumn{4}{c|}{Theoretical \ac{PL} Data}  & least squared &  Comment \\
\multicolumn{2}{c|}{Line}  & \multicolumn{2}{c|}{Polarization}  & \multicolumn{2}{c|}{Shifted Line}  & \multicolumn{2}{c|}{Polarization} & distance \\
(nm) & (eV) & perp (\%) & para (\%) & (nm) & (eV) & perp (\%) & para (\%) & Eq.~\eqref{eq:lsd}   \\
\hline
858.7	&	1.444	&	100	&	0	& & & & & &	V1'	\\
861.4	&	1.440	&	0	&	100	& 861.1 & 1.440 & 0 & 100 & & V1	\\
862.2	&	1.438	&	24	&	76  & 860.5 & 1.441 & 23 & 77 & $8.6 \cdot 10^{-5}$ &	$\mathbf{h^{a''}_{6.9}}$	\\
862.7	&	1.437	&	0	&	100	& & & & &		\\
862.9	&	1.437	&	0	&	100	& & & & &		\\
863.2	&	1.437	&	10	&	90  & 862.3 & 1.438 & 11 & 89 & $5.0 \cdot 10^{-4}$ &	$\mathbf{h^{b'''}_{7.8}}$	\\
863.6	&	1.436	&	100	&	0	& 846.4 & 1.465 & 100 & 0 & $8.5 \cdot 10^{-4}$ & $\mathbf{h^{a}_{5.4}}$	\\
864.1	&	1.435	&	48	&	52  & 975.6 & 1.271 & 47 & 53 &  $2.7 \cdot 10^{-2}$ &	$\mathbf{k^{p}_{6.2}}$	\\
864.3	&	1.435	&	0	&	100   & 864.1 & 1.435 & 0 & 100 	& $1.0 \cdot 10^{-5}$ & $\mathbf{h^{b''}_{5.4}}$\\
864.8	&	1.434	&	100	&	0  & *880.0\textcolor{white}{*} & *1.409\textcolor{white}{*} & 99 & 1 & $8.2 \cdot 10^{-4}$ &	$\mathbf{h^{p}_{9.3}}$  \\ 
865.6	&	1.433	&	0	&	100  & 864.1 & 1.435 & 0 & 100	& $1.6 \cdot 10^{-5}$ &	$h^{b''}_{5.4}$	\\
865.9	&	1.432	&	0	&	100  & 864.1 & 1.435 & 0 & 100	&$1.9 \cdot 10^{-5}$ & $h^{b''}_{5.4}$	\\
866.6	&	1.431	&	0	&	100  & 869.6 & 1.426 & 0 & 100	& $2.4 \cdot 10^{-5}$& $h^{p}_{5.4}\textrm{I}$ \\
867.1	&	1.430	&	77	&	23	 & & & & &	\\
869.6	&	1.426	&	0	&	100   & 869.6 & 1.426 & 0 & 100	& $1.3 \cdot 10^{-7}$ & $\mathbf{h^{p}_{5.4}\textrm{I}}$	\\
873.4	&	1.420	&	0	&	100   & 869.6 & 1.426 & 0 & 100	& $3.9 \cdot 10^{-5}$ & $h^{p}_{5.4}\textrm{I}$	\\
874.9	&	1.417	&	100	&	0	& 883.2 & 1.404 & 100 & 0	& $1.8 \cdot 10^{-4}$ &	$h^{b}_{3.1}$	\\
875.9	&	1.416	&	100	&	0 & 883.2 & 1.404 & 100 & 0	& $1.4 \cdot 10^{-4}$ &	$h^{b}_{3.1}$	\\
877.7	&	1.413	&	71	&	29	& & & & &	\\
880.0	&	1.409	&	0	&	100  & 881.9 & 1.406 & 0 & 100	& $9.6 \cdot 10^{-6}$ & $\mathbf{k^{a''}_{5.0}}$	\\
882.5	&	1.405	&	100	&	0   & 883.2 & 1.404 & 100 & 0	& $2.0 \cdot 10^{-6}$ &	$\mathbf{h^{b}_{3.1}}$	\\
910.7	&	1.362	&	0	&	100  & 913.8 & 1.357 & 0 & 100	& $2.1 \cdot 10^{-5}$ & $k^{b'''}_{7.8}$	\\
911.2	&	1.361	&	0	&	100   & 913.8 & 1.357 & 0 & 100	& $1.5 \cdot 10^{-5}$ & $k^{b'''}_{7.8}$ \\
913.0	&	1.358	&	0	&	100  & 913.8 & 1.357 & 0 & 100	& $1.4 \cdot 10^{-6}$ &	$k^{b'''}_{7.8}$ \\
913.4	&	1.358	&	0	&	100	 & 913.8 & 1.357 & 0 & 100 & $3.3 \cdot 10^{-7}$ &	$\mathbf{k^{b'''}_{7.8}}$  \\
914.6	&	1.356	&	0	&	100 & 913.8 & 1.357 & 0 & 100 & $1.5 \cdot 10^{-6}$ & 	$k^{b'''}_{7.8}$ \\
915.3	&	1.355	&	0	&	100  & 915.8 & 1.354 & 0 & 100 & $5.6 \cdot 10^{-7}$ & $\mathbf{k^{a}_{4.4}}$ \\
916.3	&	1.353	&	0	&	100  & 916.5 & 1.353 & 0 & 100	& &	V2	\\
926.1	&	1.339	&	0	&	100	& 929.5 & 1.334 & 0 & 100	& $3.3 \cdot 10^{-5}$  & $k^{b'''}_{8.9}$	\\
929.8	&	1.334	&	0	&	100   & 929.5 & 1.334 & 0 & 100	& $8.7 \cdot 10^{-6}$  & $\mathbf{k^{b'''}_{8.9}}$	\\
\end{tabular}
\end{ruledtabular}
\label{tab:id}
\footnotesize{$^*$PBE results shifted with 195 meV to align theoretical and experimental values for the isolated silicon vacancy.}\\
\end{table}

\subsection{Outlook}

For the modified vacancies with the most distant carbon antisites, the ZPLs are close to the isolated silicon vacancy.
Here, it is difficult to identify the individual lines.
Hence, one can consider the linewidth broadening in addition to the ZPL position.
There is a difference in linewidth of the V2 centers depending if the implantation is done with He$^+$ or Si$^{2+}$, 0.3 nm compared with 1 nm~\cite{Babin2022}. 
However, it is difficult to know which effect (surface effect, strain, or other defects) contributes toward broadening.
When comparing these experimental results to the molecular dynamics simulations done in Ref.~\onlinecite{MD2silvac2021}, one sees that He goes deeper than Si (cf. Fig.~4 in Ref.~\onlinecite{MD2silvac2021}).
If one disregards surface effects and strain, more modified vacancies could be created when the implantation is carried out with silicon ions compared to helium.
Many modified vacancy configurations (see Figure~\ref{fig:zpl}b) may contribute to increased linewidth.
However, they should anneal out at around 800 $^\circ$C, and one would expect to see decreased linewidth after such annealing.
Such a decrease in linewidth is not clearly observed. 
A distant carbon antisite could account for the defects seen in Ref.~\onlinecite{hunter2019}.
One can see carbon antisites in the molecular dynamics simulations done in Ref.~\onlinecite{MD2silvac2021}.
Even after annealing, there are carbon antisites left (cf. Fig.~8 in Ref.~\onlinecite{MD2silvac2021}).
A possible way to verify if the modified vacancies contribute to the broadening of the isolated vacancy ZPL is to map the positions of the carbon-13 around defects with, for example, the method presented in Ref.~\onlinecite{Abobeih2019}.
Using this method for an enriched sample, one could look for carbon-13 antisites that could affect the silicon vacancy.

Like modifying silicon vacancy with a carbon antisite, one can imagine modifying other defects, such as divacancy.
When the modified silicon vacancy anneals out, one possibility is that they transform to modified divacancy by interacting with a mobile carbon vacancy.
There are already additional lines (PL5-6) next to the divacancy which have been assigned to stacking fault~\cite{stackingpaper}.
If there are more lines, modified divacancy is a good starting guess.

The silicon vacancy has been suggested as a qubit candidate.
It has unique properties compared to the NV center in diamond~\cite{baranov2011silicon}, for example, the Si vacancy has a much smaller ZFS, in the MHz range.
Due to this low ZFS, an external magnetic field is required to split the ground state~\cite{Riedel2012Resonant}.
The closest modified vacancies have the same properties as the isolated silicon vacancy and should also be suitable qubit candidates.
In addition, they have an order of magnitude larger ZFS that allows operation at lower external magnetic fields.
The lower symmetry of the modified vacancies does not have to be seen as a drawback.
The low symmetry configuration of the divacancy has shown promise for quantum technologies~\cite{Miao2019,Miao2020}.

\section{Conclusion}

In this paper, we have searched for a defect that explains the experimental signals in the vicinity of silicon vacancy in a defect database generated by ADAQ containing 8355 single and double intrinsic defects.
Based on this search, the modified silicon vacancy with a carbon antisite is the only defect candidate.
The modified vacancy has similar properties as the silicon vacancy, and the carbon antisite is a small enough perturbation that does not change the spin or charge state of the silicon vacancy.
This defect has several configurations which have been identified with the new PL experiments performed in this paper.
Foremost, the $k^{a''}_{5.0}$ configuration has been identified to be the source of the $R_1$ EPR signal and 880.0 nm PL line.
Moreover, the $h^{b}_{3.1}$ configuration, which is one of the most stable and closest configurations, is most likely responsible for the 875.9 nm PL line.
Other configurations explain lines closer to the isolated vacancy and might explain line broadening as well.
However, for these lines, further studies and experiments are needed to rule out other effects, such as strain, before any conclusion can be drawn.
Finally, our work demonstrates a high-throughput approach to search for point defects that explain experimental results is a promising direction forward.

\section{Methods}
\label{sec:method}

\subsection{Computational Details}

The computations were carried out with Vienna Ab initio Simulation Package (VASP)~\cite{VASP,VASP2} (v. 5.4.4), which, in turn, uses a plane wave basis set and the projector augmented wave (PAW)~\cite{PAW,Kresse99} method.
For the manual calculations, we employ the semi-local exchange-correlation functional of Perdew, Burke, and Erzenerhof (PBE)~\cite{PBE} and the screened hybrid functional of Heyd, Scuseria, and Ernzerhof (HSE06)~\cite{HSE03,HSE06} with the mixing parameter $\alpha$ set to the standard value (25\%).
For the PBE calculations, the plane wave energy and kinetic energy cutoff are set to 600 and 900 eV, respectively, which are the settings used in \ac{ADAQ}.
Whereas, for the HSE calculations, these are reduced to 420 and 840 eV, respectively.
The total energy criterion is set to $10^{-6}$ eV for PBE, and $10^{-4}$ eV for HSE.
The structural minimization criterion is set to $5 \times 10^{-5}$ eV for PBE, and $10^{-2}$ eV/\AA\ for HSE.
The pseudopotentials for C is labeled 05jan2001 and for Si is labeled 08april2002.
Two supercell sizes have been employed with 576 (6x6x2) and 2304 (12x12x2) atoms, respectively.
In this paper, only $\Gamma$-point sampling of the Brillouin zone with Gaussian smearing is used, and $\Psi_k=\Psi_{-k}^*$ is the only symmetry used.

The following defect properties~\cite{ADAQ} are calculated in this paper:
\begin{itemize}
	\item Formation energy~\cite{vanderwalle} for the charged defect:
		\begin{equation}
			\Delta H_D^q(E_f,\mu) = [E_{D,q}-E_H] - \sum_i{n_i\mu_i} + qE_f + E_{\mathrm{corr}}(q).
			\label{eq:form}
		\end{equation}
			Where $E_{D,q}$ and $E_H$ are the total energies of the charged defect supercell and neutral host supercell, $\mu_i$ are the chemical potentials (only the rich limits are used: stoichiometric condition), $n_i$ are the added or missing elements, $q$ is the charge, and $E_f$ is the Fermi energy.
			For the charge correction term $E_{\mathrm{corr}}(q)$, the Lany-Zunger correction is used~\cite{LanyZunger08}.
	\item The binding energy $H^b$ is defined as:
		\begin{equation}
			H^b = \Delta H_{D,q}(E_f,\mu)[A] + \Delta H_{D,q}(E_f,\mu)[B] - \Delta H_{D,q}(E_f,\mu)[AB].
			\label{eq:bind}
		\end{equation}
	 Here A and B are the separate components and AB is the cluster.
	 Defects with a positive binding energy are considered to be stable.
	 For the modified vacancy, this equation becomes:
	 	\begin{equation}
			H^b = \Delta H_{D,-}(E_f,\mu)[V_{Si}] + \Delta H_{D,0}(E_f,\mu)[C_{Si}] - \Delta H_{D,-}(E_f,\mu)[V_{Si}-C_{Si}].
			\label{eq:bind2}
		\end{equation}
	 For a given modified vacancy (AB), the reference values for components A and B are taken from the corresponding isolated defects.
	 For example, a modified vacancy with a silicon vacancy at $h$ and carbon antisite at $k$ uses the corresponding energies for these defects in the isolated case.
	\item \ac{ZPL} is calculated as the total energy difference between the ground and excited state~\cite{Gali:PRL2009}.
	The excited state is found by using constrained DFT ($\Delta$-SCF method)~\cite{cDFT,GACDFT}.
	The relative \ac{ZPL}s can be compared within an accuracy of 100 meV~\cite{methodology_paper}.
	The lowest energy optical excitation is from the occupied $a_1$ state to the unoccupied $a_1$ state in one spin channel~\cite{ivady2018first}.
	\item \ac{TDM} between the ground and excited state is calculated using the wave functions from the relaxed ground and excited state, which provide an accurate polarization~\cite{davidsson2020theoretical}.
	\item \ac{ZFS} is approximated using the dipole-dipole interaction of the spins~\cite{Ivady2014,ivady2018first} and is calculated with the implementation in VASP.
	From the \textbf{D}-tensor, the D value is expressed as $D= 3D_{zz}/2$ and the E splitting as $E=(D_{yy}-D_{xx})/2$
\end{itemize}

\subsection{Samples and Experimental Details}
\label{sec:ex_method}

All the samples are cut from the same wafer and irradiated to 17 cm-2 with electrons of energy 2 MeV.
The samples are considered to have negligible strain due to the good PL agreement with the isolated silicon vacancy and divacancy.
Two of the samples were annealed at different temperatures for half an hour from 200 $^{\circ}$C to 800 $^{\circ}$C for investigating the behavior in the range of the silicon vacancy zero phonon lines at different temperatures (Due to insignificant differences only 400 $^{\circ}$C and 800 $^{\circ}$C samples are shown in Figure~\ref{fig:exp}).
All the PL measurements were performed with a Jobin Yvon HR460 monochromator.
The monochromator is equipped with 1200 grooves/mm grating blazed at 750 nm and an InGaAs multi-channel detector, ensuring resolution $\approx$1 Å and optimum sensitivity in the silicon-vacancy range (wavelengths $\approx$ 858-980 nm).
All the samples are mounted in a variable temperature closed-cycle cryostat and cooled down to 3.5 K.
We use a 730 nm diode laser for exciting V1 and 890 nm from tuneable Ti-sapphire laser as an excitation laser for V2 to eliminate the background luminescence from V1.

\section*{Acknowledgments}

We acknowledge support from the Knut and Alice Wallenberg Foundation through WBSQD2 project (Grant No. 2018.0071).
Support from the Swedish Government Strategic Research Area Swedish e-science Research Centre (SeRC) and the Swedish Government Strategic Research Area in Materials Science on Functional Materials at Linköping University (Faculty Grant SFO-Mat-LiU No. 2009 00971) are gratefully acknowledged.
JD acknowledges support from the Covid-19 SeRC transition grant.
RA acknowledges support from the Swedish Research Council (VR) Grant No. 2020-05402.
The computations were enabled by resources provided by the Swedish National Infrastructure for Computing (SNIC) at NSC partially funded by the Swedish Research Council through grant agreement no. 2018-05973.

\section*{Contributions}

J.D. conceived the project, suggested the defect to study by searching through the database, and wrote the majority of the paper.
R.B. ran the manual calculations and produced all figures and tables related to the computational results in the paper.
D.S. and I.G.I. performed the \ac{PL} measurements and produced the experimental figure and table.
J.D., R.B., and V.I. interpreted the results and identified the configurations.
V.I., R.A., and I.A.A. supervised the work.
All authors modified and discussed the paper together.

\section*{Competing Interests}
The authors declare no competing interests.

\bibliographystyle{apsrev4-1}
\bibliography{../../references}

\end{document}